# Performance Analysis of RIS-Based nT-FSO Link Over G-G Turbulence With Pointing Errors

Alain R. Ndjiongue, *Senior Member, IEEE*, Telex M. N. Ngatched, *Senior Member, IEEE*, Octavia A. Dobre, *Fellow, IEEE*, Ana G. Armada, *Senior Member, IEEE*, and Harald Haas, *Fellow, IEEE*

*Abstract*—One of the main problems faced by communication systems is the presence of skip-zones in the targeted areas. With the deployment of the fifth-generation mobile network, solutions are proposed to solve the signal loss due to obstruction by buildings, mountains, and atmospheric or weather conditions. Among these solutions, re-configurable intelligent surfaces (RIS), which are newly proposed modules, may be exploited to reflect the incident signal in the direction of dead zones, increase communication coverage, and make the channel smarter and controllable. In this paper, we tackle the skip-zone problem in near-terrestrial free-space optical (nT-FSO) systems using RIS. We carry out a performance analysis of RIS-aided nT-FSO links affected by turbulence and pointing errors, for both heterodyne detection (HD) and intensity modulation-direct detection (IM/DD) techniques. Turbulence is modeled using the Gamma-Gamma (G-G) distribution. We analyze the model and provide exact closed-form expressions of the probability density function, cumulative distribution function, and moment generating function of the end-to-end signal-to-noise ratio, $\gamma$. Capitalizing on these statistics, we evaluate the system performance through the outage probability, $\overline{P}_{out}$, ergodic channel capacity, $\overline{C}$, and average bit-error-rate, $\overline{P}_b$, for selected binary modulation schemes. Numerical results obtained for different turbulence levels and pointing errors confirm that the HD technique outperforms IM/DD even in RIS-aided nT-FSO systems. These results also show that using a blue color offers better channel capacity and communication performance compared to red and green colors.

## I. INTRODUCTION

The recent extensive investigation of optical wireless communications in the outdoor environment, also called free-space optical (FSO), is motivated by its advantages compared to its radio frequency (RF) counterpart, especially in point-to-point networks. These advantages include larger bandwidth, higher channel capacity, and cost-effectiveness due to an unlicensed environment [1], which can be leveraged to solve the bandwidth limitation in the RF technology. Its most prominent applications are satellite-to-ground, satellite-to-satellite, and near-terrestrial FSO (nT-FSO) systems such as building-to-building (B2B) communications. Besides turbulence, pointing

Alain. R. Ndjiongue, T. M. N. Ngatched, and O. A. Dobre are with the Faculty of Engineering and Applied Science, Memorial University of Newfoundland, Canada.
Ana G. Armada is with the Signal Theory and Communications Department, Universidad Carlos III de Madrid, Spain.
Harald Haas is with the LiFi Research and Development Center, Department of Electronic and Electrical Engineering, the University of Strathclyde, Glasgow, United Kingdom.

errors, and attenuation that affect optical signals over the FSO channel, signal obstruction due to buildings or trees can prevent the transmitted message to reach the destination. We attempt to solve this obstruction's problem in nT-FSO systems, affected by moderate-to-strong turbulence levels and pointing errors, using re-configurable intelligent surfaces (RIS). RIS are electromagnetic devices with electronically controllable characteristics. They can reflect, refract, extinct, or scatter the incoming signal with an impact on its amplitude, phase, and polarization. The design of RIS modules depends on the application.

The RIS module is a planar array of multiple mirrors used to guide the incoming signal toward a targeted area and re-configure the transmission channel [2]. It offers wireless networks several advantages over competing technologies such as relay systems. In addition to their low power consumption, the RIS module is made of electronically controllable elements. These advantages have recently triggered intensive investigations of the technology. It has lately been proposed to solve the dead zone problems in RF networks and create smart communication channels and environments [3]–[5], or serve as a wave-guide in visible light communications [6]. To create a smart RF channel, authors in [3] investigated principle, challenges, and opportunities related to the use of RIS modules in an indoor environment. Most works in RIS-aided communication networks focus on simplex systems. The authors in [4] analyzed a two-way communication system assisted by RIS modules. The RIS influence on spatial modulations is investigated in [7], where the authors proposed a low-complexity and fast antenna selection algorithm, while in [5], the authors introduced RIS-space shift keying and RIS-spatial modulation schemes. As part of the channel, the RIS elements may decisively impact wireless communication systems' performance, leading to the need for new pre-coding designs [8]. The RIS concept can be extended to re-configurable optical components. For example, in [9], the authors proposed re-configurable photo-detectors and explored the use of blind interference alignment to achieve a multiplexing gain without cooperation among light-emitting diodes.

Due to the presence and locations of obstacles, the use of RIS in FSO are suitable for nT-FSO communication systems such as B2B[1]. Investigation on using RIS in nT-FSO systems is still in its infancy; however, it is predicted that it will attract significant research interests. Early work on using a RIS module in nT-FSO systems is proposed in [10], [11]. In [10], the authors discuss the implementation of a RIS-based FSO system considering controllable multi-branches. They neglect turbulence over the channels because of the length shortness of branches, which was less than 500 m, allowing the analysis to consider only pointing errors. In [11], based on the central limit theorem, the authors exploit the Gaussian distribution to approximate a Gamma-Gamma (G-G) channel for a large number of transmitting signals. In contrast with [10] and [11], this paper considers a RIS-based nT-FSO communication system exploiting a single light-ray to transmit data over a G-G channel with pointing errors.

---

[1] A B2B environment is an nT-FSO data transmission environment where the information is transferred between buildings.

To the best of our knowledge, using RIS in an nT-FSO system characterized by G-G turbulence with pointing errors, has not yet been proposed in the open literature and represents the motivation of this paper. The main goals of the paper are to show the numerical analysis and performance of the considered nT-FSO system. To this end, we make the following contributions: (*i*) we derive closed-form unified statistical expressions of the probability density function (PDF), cumulative distribution function (CDF), and moment generating function (MGF) of the end-to-end signal-to-noise ratio (SNR), $\gamma$; (*ii*) based on these results, we derive the outage probability (OP), $P_{out}$, the average ergodic channel capacity, $\overline{C}$, and the average bit-error-rate (BER), $\overline{P}_b$, for selected binary modulation schemes including coherent binary frequency-shift keying (CBFSK), non-coherent binary frequency-shift keying (NBFSK), coherent binary phase-shift keying (CBPSK), and differential binary phase-shift keying (DBPSK); (*iii*) next, we derive closed-form expressions of the diversity order and coding gain for the proposed RIS-based nT-FSO; (*iv*) finally, we present numerical results for different turbulence and pointing error levels, and compare the performance between transmissions using the red, green, and blue light colors.

## II. SYSTEM AND CHANNEL MODELS

### A. System Model

The environment under study is a cascaded system of a single light ray traveling from source (S) to destination (D) after reflection on a RIS element, as shown in Fig. 1. There is no direct link between S and D owing to obstructions. The RIS module, located at the top of a building, serves as a reflector to the incoming signal and ensures that the transmitted light points to the receiver. We assume that both channel portions, which will be denoted as the system sub-channels, exhibit moderate-to-strong turbulence levels, and the light intensity over them undergoes the same attenuation level. It is also assumed that, at each receiving end, the detectors face similar pointing errors.

### B. The FSO Channel

The FSO link is subject to three main signal impairment factors: pointing errors, atmospheric turbulence, and attenuation. These impairment sources, each in its way, affect the transmitted optical signal, $I$, which can be expressed as $I = I_p I_a I_l$, where $I_p$, $I_a$, and $I_l$ represent the received intensity affected by pointing errors, atmospheric turbulence,

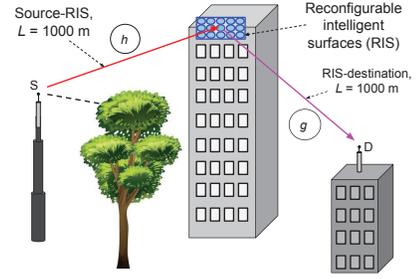

Fig. 1: A model of RIS-aided nT-FSO system.

and attenuation, respectively. The PDF describing pointing errors, $f_p(I_p)$, is given by [1, Eq. (3)], [12]

$$f_p(I_p) = \frac{\zeta^2}{A_o^{\zeta^2}} I_p^{\zeta^2-1}, \quad 0 \leq I_p \leq A_0, \quad (1)$$

where $\zeta$ is the ratio of equivalent beam radius at the receiver and pointing displacement standard deviation, $A_o = [\text{erf}(v)]^2$, erf($\cdot$) being the error function and $v = r\sqrt{\pi}/\sqrt{2}w_z$. Here, $r$ and $w_z$ are the radius of the receiver aperture and the beam waist, respectively [1], [12]. In general, the G-G distribution is used to model channels characterized by moderated-to-strong turbulence levels. Its PDF is given by [11]

$$f_{I_a}(I_a) = \frac{2(\alpha\beta)^{\frac{\alpha+\beta}{2}}}{\Gamma(\alpha)\Gamma(\beta)} I_a^{\frac{\alpha+\beta}{2}-1} K_{\alpha-\beta}\left(2\sqrt{\alpha\beta I_a}\right), \quad (2)$$

where $K_j(\cdot)$ is the $j^{th}$-order modified Bessel function of second kind. The values of $\alpha$ and $\beta$ can be calculated using the Rytov variance, $\sigma_2^2 = 0.492 C_n^2 \eta^{\frac{7}{6}} L^{\frac{11}{6}}$, which depends on the altitude-dependent index, $C_n^2$, characterized by the transmission environment, the angular wavenumber, $\eta = \frac{2\pi}{\lambda}$, the aperture diameter, $D$, and the transmission distance, $L$. They are respectively given by [13]

$$\alpha = \left[\exp\left(\frac{0.49\sigma_2^2}{\left(1 + 0.18d^2 + 0.56\sigma_2^{12/5}\right)^{7/6}}\right) - 1\right]^{-1}, \quad (3)$$

and

$$\beta = \left[\exp\left(\frac{0.51\sigma_2^2 \left(1 + 0.69\sigma_2^{12/5}\right)^{-5/6}}{\left(1 + 0.9d^2 + 0.62\sigma_2^{12/5}\right)^{5/6}}\right) - 1\right]^{-1}, \quad (4)$$

where $d = \sqrt{\eta D^2/4L}$. Finally, the path loss, which is considered as constant for a given weather condition and link distance [1], is given by the Beer-Lambert law as $I_l = e^{-\delta L}$, where $\delta$ is the attenuation factor.

## III. CLOSED-FORM STATISTICAL ANALYSIS

In this section, we derive unified closed-form expressions for the PDF, CDF, and MGF of the end-to-end SNR.

### A. End-to-End Signal-to-Noise Ratio (SNR)

We assume that the RIS module plays only a reflective function and does not allow light through. We also assume perfect knowledge of the channel phases at RIS and destination. The detected signal can be expressed as $y = \sqrt{E_s}(h\mu e^{j\theta}g)x + n$, where $E_S$ is the symbol energy, $h$ and $g$ are respectively the S-RIS and RIS-D complex channel vectors, $\mu e^{j\theta}$ characterizes the RIS element with $\mu$ being its amplitude reflection coefficient and $\theta$ its induced phase [5], [14]. $x$ and $y$ are the transmitted and received symbols, respectively, and $n$ is the additive white Gaussian noise at the destination. At RIS, the main goal is to reflect the original signal in such a way to optimize signal reception at D, which can be done through end-to-end SNR maximization. This SNR is defined by $\gamma = \bar{\gamma}|h\mu e^{j\theta}g|^2$, where $\bar{\gamma} = E_s/N_0$ represents the average SNR in both S-RIS and RIS-D sub-channels, and $N_0$ is the noise power spectral density at D.

### B. PDF of the End-to-End SNR

The overall system's gain is given by $h\mu e^{j\theta}g$, where the quantity $\mu e^{j\theta}$ is deterministic in contrast to $h$ and $g$, which are random variables. Thus, the SNR's PDF, $f_\gamma(\gamma)$, can be calculated from the SNRs, $\gamma_h$ and $\gamma_g$, and can be evaluated as [15, Eq. (5)]

$$f_\gamma(\gamma) = \int_0^\infty f_{\gamma_h}(t) f_{\gamma_g}\left(\frac{\gamma}{t}\right) \frac{1}{t} dt, \quad (5)$$

where $f_{\gamma_h}(\cdot)$ and $f_{\gamma_g}(\cdot)$ are respectively the PDFs of the S-RIS and RIS-D sub-channel's SNRs, $\gamma_h$ and $\gamma_g$. With the assumption of a constant weather condition over the environment, both parts of the channel can be modeled by a combined distribution including pointing errors and turbulence levels. The resulting unified PDF, $f_{\gamma_i}(\gamma_i)$ is expressed as [1, Eq. (7)]

$$f_{\gamma_i}(\gamma_i) = \frac{M}{\gamma_i} \mathbf{G}_{1,3}^{3,0}\left[Q\left(\frac{\gamma_i}{\bar{\gamma}_i}\right)^{\frac{1}{a}} \middle| \begin{array}{c} \zeta^2+1 \\ \zeta^2, \alpha, \beta \end{array}\right], \quad (6)$$

where $i \in \{h, g\}$, $M = \frac{\zeta^2}{a\Gamma(\alpha)\Gamma(\beta)}$, $Q = \frac{\zeta^2\alpha\beta}{(1+\zeta^2)}$, $a \in \{1, 2\}$ indicates whether the transmission exploits the HD ($a = 1$) or IM/DD ($a = 2$) techniques [1], and $\mathbf{G}_{p,q}^{m,n}\left[z\middle|\begin{array}{c}a_p\\b_q\end{array}\right]$ is the Meijer-G function. We sequentially substitute $\gamma_i$ by $t$ and $\frac{\gamma}{t}$ in Eq. (6), and obtain $f_{\gamma_h}(t)$ and $f_{\gamma_g}\left(\frac{\gamma}{t}\right)$ respectively as

$$f_{\gamma_h}(t) = \frac{M}{t} \mathbf{G}_{1,3}^{3,0}\left[Q\left(\frac{t}{\bar{\gamma}_h}\right)^{\frac{1}{a}} \middle| \begin{array}{c} \zeta^2+1 \\ \zeta^2, \alpha, \beta \end{array}\right], \quad (7)$$

and

$$f_{\gamma_g}\left(\frac{\gamma}{t}\right) = \frac{Mt}{\gamma} \mathbf{G}_{1,3}^{3,0}\left[Q\left(\frac{\gamma}{\bar{\gamma}_g t}\right)^{\frac{1}{a}} \middle| \begin{array}{c} \zeta^2+1 \\ \zeta^2, \alpha, \beta \end{array}\right], \quad (8)$$

where $\bar{\gamma}_h$ and $\bar{\gamma}_g$ are average values of the SNRs $\gamma_h$ and $\gamma_g$, respectively. In Eq. (8), the variable $t$ appears at the denominator. To obtain a Meijer-G function with a numerator-based variable $t$, we apply the reflection property of the Meijer-G function, given by [16]

$$\mathbf{G}_{p,q}^{m,n}\left[z\middle|\begin{array}{c}A_p\\B_q\end{array}\right] = \mathbf{G}_{q,p}^{n,m}\left[z^{-1}\middle|\begin{array}{c}1-B_q\\1-A_p\end{array}\right], \quad (9)$$

to Eq. (8) and obtain

$$f_{\gamma_g}\left(\frac{\gamma}{t}\right) = \frac{Mt}{\gamma} \mathbf{G}_{3,1}^{0,3}\left[\frac{1}{Q}\left(\frac{\bar{\gamma}_g}{\gamma}\right)^{\frac{1}{a}} t^{\frac{1}{a}} \middle| \begin{array}{c} 1-\zeta^2, 1-\alpha, 1-\beta \\ -\zeta^2 \end{array}\right]. \quad (10)$$

To get the end-to-end SNR's PDF, $f_\gamma(\gamma)$, we substitute Eqs. (7) and (10) into Eq. (5), which leads to

$$f_\gamma(\gamma) = \frac{M^2}{\gamma} \int_0^\infty \frac{1}{t} \mathbf{G}_{1,3}^{3,0}\left[Q\left(\frac{t}{\bar{\gamma}_h}\right)^{\frac{1}{a}} \middle| \begin{array}{c} \zeta^2+1 \\ \zeta^2, \alpha, \beta \end{array}\right]$$
$$\times \mathbf{G}_{3,1}^{0,3}\left[\frac{1}{Q}\left(\frac{\bar{\gamma}_g}{\gamma}\right)^{\frac{1}{a}} t^{\frac{1}{a}} \middle| \begin{array}{c} 1-\zeta^2, 1-\alpha, 1-\beta \\ -\zeta^2 \end{array}\right] d\gamma. \quad (11)$$

Applying the change of variable $X = t^{\frac{1}{a}} \Rightarrow t = X^a$, and $dt = aX^{a-1}dX$, we obtain

$$f_\gamma(\gamma) = \frac{aM^2}{\gamma} \int_0^\infty \frac{1}{X} \mathbf{G}_{1,3}^{3,0}\left[\frac{QX}{\bar{\gamma}_h^{\frac{1}{a}}} \middle| \begin{array}{c} \zeta^2+1 \\ \zeta^2, \alpha, \beta \end{array}\right]$$
$$\times \mathbf{G}_{3,1}^{0,3}\left[\frac{1}{Q}\left(\frac{\bar{\gamma}_g}{\gamma}\right)^{\frac{1}{a}} X \middle| \begin{array}{c} 1-\zeta^2, 1-\alpha, 1-\beta \\ -\zeta^2 \end{array}\right] dX. \quad (12)$$

With the help of [17, Eq. (07.34.21.0011.01)], we solve the integral in Eq. (12), apply the identity in Eq. (9), and obtain the exact unified PDF of end-to-end SNR, $f_\gamma(\gamma)$, as

$$f_\gamma(\gamma) = \frac{aM^2}{\gamma} \mathbf{G}_{2,6}^{6,0}\left[Q^2\left(\frac{\gamma}{\bar{\gamma}}\right)^{\frac{1}{a}} \middle| \begin{array}{c} \zeta^2+1, \zeta^2+1 \\ \zeta^2, \alpha, \beta, \zeta^2, \alpha, \beta \end{array}\right], \quad (13)$$

where $\bar{\gamma} = \bar{\gamma}_g \bar{\gamma}_h$.

### C. CDF of the End-to-End SNR

The CDF of the end-to-end SNR, $F_\gamma(\gamma)$, can be calculated as $F_\gamma(\gamma) = \int_0^\infty f_\gamma(\gamma) d\gamma$. Substituting the expression of $f_\gamma(\gamma)$ (Eq. (13)), and permuting the variables $\gamma$, $x$, and $\infty$, we obtain

$$F_\gamma(\gamma) = aM^2 \int_0^\gamma \frac{1}{x} \mathbf{G}_{2,6}^{6,0}\left[Q^2\left(\frac{x}{\bar{\gamma}}\right)^{\frac{1}{a}} \middle| \begin{array}{c} \zeta^2+1, \zeta^2+1 \\ \zeta^2, \alpha, \beta, \zeta^2, \alpha, \beta \end{array}\right] dx. \quad (14)$$

With the help of [17, Eq. (07.34.21.0084.01)], we solve the integral in Eq. (14), to get the closed-form expression of $F_\gamma(\gamma)$ as

$$F_\gamma(\gamma) = M_0 \mathbf{G}_{2a+1,6a+1}^{6a,1}\left[Q_0\left(\frac{\gamma}{\bar{\gamma}}\right) \middle| \begin{array}{c} 1, \Delta_1 \\ \Delta_2, 0 \end{array}\right], \quad (15)$$

where $M_0 = \frac{M^2 a^{2(\alpha+\beta-1)}}{(2\pi)^{2(a-1)}}$, $Q_0 = \frac{Q^{2a}}{a^{4a}}$, $\Delta_1 = \frac{\zeta^2+1}{a}, \ldots, \frac{\zeta^2+a}{a}$, $\frac{\zeta^2+1}{a}, \ldots, \frac{\zeta^2+a}{a}$ with $2a$ terms, and $\Delta_2 = \frac{\zeta^2}{a}, \ldots, \frac{\zeta^2+a-1}{a}, \frac{\alpha}{a}, \ldots, \frac{\alpha+a-1}{a}, \frac{\beta}{a}, \ldots, \frac{\beta+a-1}{a}, \frac{\zeta^2}{a}, \ldots, \frac{\zeta^2+a-1}{a}, \frac{\alpha}{a}, \ldots, \frac{\alpha+a-1}{a}, \frac{\beta}{a}, \ldots, \frac{\beta+a-1}{a}$ with $6a$ terms.

### D. Moment Generating Function (MGF)

The MGF, $\Omega_\gamma(s)$, is readily calculated from the CDF as [1, Eq. (15)]

$$\Omega_\gamma(s) = s \int_0^\infty \exp(-\gamma s) F_\gamma(\gamma) d\gamma. \quad (16)$$

Substituting Eq. (15) into Eq. (16), we obtain

$$\Omega_\gamma(s) = M_0 s \int_0^\infty e^{-\gamma s} \mathbf{G}_{2a+1,6a+1}^{6a,1}\left[Q_0\left(\frac{\gamma}{\bar{\gamma}}\right) \middle| \begin{array}{c} 1, \Delta_1 \\ \Delta_2, 0 \end{array}\right] d\gamma. \quad (17)$$

Using [18, Eq. (7.813.1)], we solve the integral in Eq. (17) and obtain the closed-form and unified expression of the MGF as

$$\Omega_\gamma(s) = M_0 \mathbf{G}_{2a+2,6a+1}^{6a,2}\left[\frac{Q_0}{\bar{\gamma}s} \middle| \begin{array}{c} 0, 1, \Delta_1 \\ \Delta_2, 0 \end{array}\right]. \quad (18)$$

## IV. APPLICATIONS

In this section, we analyze the performance of the proposed RIS-aided nT-FSO system based on the OP, $P_{out}$, ergodic channel capacity, $\overline{C}$, and average BER, $\overline{P}_b$, for selected binary schemes.

### A. Outage Probability

Outage occurs when the end-to-end SNR, $\gamma$, falls below a threshold value, $\gamma_{th} = e^{2R-1}$, predefined for a specific quality-of-service, $R$ being the transmission rate. This implies that under such conditions, the system does not reach the specific rate $R$. The OP, $P_{out}$, which defines this failure, can be readily calculated from Eq. (15) by finding $F_\gamma(\gamma_{th})$.

### B. Ergodic Channel Capacity

In the proposed system, the channel state information is not available at the transmitter and data is transmitted without instantaneous feedback, which reduce system capacity [1]. The transmitted symbol is long enough so that data is encoded over all the possible channel fading states, and the atmospheric turbulence, which is slow-fading in nT-FSO, remains constant over the symbol transmission, combined with the effects of the pointing errors that make the signal fluctuate at a very high rate [1]. Thus, the overall channel statistical properties can be evaluated during the transmission of a single symbol. Therefore, the ergodic channel analysis can be performed [1], [19]. The ergodic channel capacity, $\overline{C}$, is given by

$$\overline{C} = \frac{1}{\ln(2)} \int_0^\infty \ln(1 + \chi\gamma) f_\gamma(\gamma) d\gamma, \quad (19)$$

where $\chi = 1$ for HD and $\chi = \frac{e}{2\pi}$ for IM/DD [1]. Exploiting the Meijer's G-function representation of $\ln(1+x)$ [17, Eq. (07.34.03.0456.01)] and substituting Eq. (13) in Eq. (19), $\overline{C}$ becomes

$$\overline{C} = \frac{aM^2}{\ln(2)} \int_0^\infty \frac{1}{\gamma} \mathbf{G}_{2,2}^{1,2}\left[\chi\gamma \middle| \begin{matrix} 1,1 \\ 1,0 \end{matrix}\right] \\ \times \mathbf{G}_{2,6}^{6,0}\left[Q^2\left(\frac{\gamma}{\overline{\gamma}}\right)^{\frac{1}{a}} \middle| \begin{matrix} \zeta^2+1, \zeta^2+1 \\ \zeta^2, \alpha, \beta, \zeta^2, \alpha, \beta \end{matrix}\right] d\gamma. \quad (20)$$

With the help of [17, Eq. (07.34.21.0013.01)], we evaluate the integral in Eq. (20) and obtain a closed-form unified expression of $\overline{C}$ as

$$\overline{C} = \frac{M_0}{\ln(2)} \mathbf{G}_{2a+2,6a+2}^{6a+2,1}\left[\frac{Q_0}{\chi\overline{\gamma}} \middle| \begin{matrix} 0, 1, \Delta_1 \\ \Delta_2, 0, 0 \end{matrix}\right]. \quad (21)$$

### C. Average Bit-Error-Rate (BER) for Selected Binary Schemes

In data transmission, the BER is a classical metric used to evaluate the system performance. Considering that in the proposed system, binary schemes are used to modulate the data before transmission, the average BER, $\overline{P}_b$, can be evaluated using [20, Eq. (13)]

$$\overline{P}_b = \frac{q^p}{2\Gamma(p)} \int_0^\infty e^{-q\gamma} \gamma^{p-1} F_\gamma(\gamma) d\gamma, \quad (22)$$

where the pair $(p, q)$ defines the binary modulation schemes [1]. The values of $p$ and $q$ for selected modulation schemes,

TABLE I: Values of $p$ and $q$ for Selected Binary Modulation Schemes.

| Scheme | $p$ | $q$ |
|---|---|---|
| CBFSK | 0.5 | 0.5 |
| NBFSK | 1 | 0.5 |
| CBPSK | 0.5 | 1 |
| DBPSK | 1 | 1 |

namely, CBFSK, NBFSK, CBPSK, and DBPSK, are provided in Table I. Substituting Eq. (15) into Eq. (22) leads to

$$\overline{P}_b = \frac{q^p M_0}{2\Gamma(p)} \int_0^\infty \frac{e^{-q\gamma}}{\gamma^{1-p}} \mathbf{G}_{2a+1,6a+1}^{6a,1}\left[Q_0\left(\frac{\gamma}{\overline{\gamma}}\right) \middle| \begin{matrix} 1, \Delta_1 \\ \Delta_2, 0 \end{matrix}\right] d\gamma. \quad (23)$$

Using [18, Eq. (7.813.1)], the integral in Eq. (23) can be solved to obtain a closed-form unified expression of the average BER, $\overline{P}_b$, as

$$\overline{P}_b = \frac{M_0}{2\Gamma(p)} \mathbf{G}_{2a+2,6a+1}^{6a,2}\left[\frac{Q_0}{q\overline{\gamma}} \middle| \begin{matrix} 1-p, 1, \Delta_1 \\ \Delta_2, 0 \end{matrix}\right]. \quad (24)$$

### D. Diversity Order and Coding Gain

The diversity order defines the increase in SNR due to some diversity schemes. Practically, it determines the slope of the $\overline{P}_b = f(\overline{\gamma})$ curve. On the other hand, the coding gain is the shift between SNR levels for coded and un-coded systems, required to reach the same $\overline{P}_b$. At high SNR, the average BER, $\overline{P}_b$, can be approximated as $\overline{P}_b = (G_c \overline{\gamma})^{-G_d}$ [21]. We use the Meijer-G function expansion [17, Eq. (07.34.06.0017.01)], associated with the unity of $\lim_{x\to\infty} {}_c F_d[e; f; x]$ [1], [22], to find the unified asymptotic expression of $\overline{P}_b$, as

$$\overline{P}_b \approx \frac{M_0}{2\Gamma(p)} \sum_{k=1}^{6a} \xi(i,j,k) \left[\frac{q\overline{\gamma}}{Q_0}\right]^{-(\Delta_{2,k})}, \quad (25)$$

where $\xi(i,j,k)$ is expressed as

$$\xi(i,j,k) = \frac{\Gamma(\Delta_{2,k}+p) \prod_{j=1; j\neq k}^{6a} \Gamma(\Delta_{2,j} - \Delta_{2,k})}{\Delta_{2,k} \prod_{i=3}^{2a+2} \Gamma(\Delta_{1,i} - \Delta_{2,k})}, \quad (26)$$

where $\Delta_{1,i} = \Delta_{1,1}, \Delta_{1,2}, \ldots, \Delta_{1,2a+2}$ with $2a+2$ terms, $\Delta_{2,j} = \Delta_{2,1}, \Delta_{2,2}, \ldots, \Delta_{2,6a}$ with $6a$ terms, and $\Delta_{2,k} = \Delta_{2,1}, \Delta_{2,2}, \ldots, \Delta_{2,6a}$ with $6a$ terms. By comparing Eq. (25) to $\overline{P}_b = (G_c \overline{\gamma})^{-G_d}$, we obtain the unified expressions of the diversity order and coding gain as $G_d = \min\left(\frac{\zeta^2}{a}, \ldots, \frac{\zeta^2+a-1}{a}, \frac{\alpha}{a}, \ldots, \frac{\alpha+a-1}{a}, \frac{\beta}{a}, \ldots, \frac{\beta+a-1}{a}\right)$ and

$$G_c = \frac{q}{Q_0} \left[\sum_{k=1}^{6a} \xi(i,j,k) \frac{M_0}{2\Gamma(p)}\right]^{\frac{-1}{\Delta_{2,k}}},$$

respectively. Note that $G_d$ will be found between $\zeta^2$, $\alpha$, and $\beta$ for the HD technique, and between $\frac{\zeta^2}{2}$, $\frac{\zeta^2+1}{2}$, $\frac{\alpha}{a}$, $\frac{\alpha+1}{2}$, $\frac{\beta}{2}$, $\frac{\beta+1}{2}$ for the IM/DD technique.

TABLE II: Values of $\alpha$ and $\beta$ used in the analysis. $L = 1$ km and $D = 1$ mm.

| Color | Red | Blue | Green |
|---|---|---|---|
| $\lambda$ | 700 | 470 | 530 |
| $C_n^2 = 2 \times 10^{-11}$ | | | |
| $\alpha$ | 10.9537 | 12.5331 | 13.2818 |
| $\beta$ | 2.9833 | 4.6787 | 5.7795 |
| $C_n^2 = 3 \times 10^{-12}$ | | | |
| $\alpha$ | 4.9477 | 5.6690 | 6.0130 |
| $\beta$ | 1.2310 | 1.4315 | 1.5682 |
| $C_n^2 = 5 \times 10^{-14}$ | | | |
| $\alpha$ | 2.9428 | 2.5012 | 2.3664 |
| $\beta$ | 2.5605 | 2.0807 | 1.9221 |

## V. NUMERICAL RESULTS

We consider a nT-FSO transmission environment in which S and D are situated at the same distance, $L = 1$ km, from the RIS (see Fig. 1). Three transmitters, each equipped with a different colored light source, are available to be used. The corresponding wavelengths are $\lambda_r = 700$ nm, $\lambda_g = 530$ nm, and $\lambda_b = 470$ nm, for the red, green, and blue light, respectively. The two system's sub-channels are characterized by the same refractive structure and index, $C_n^2$, which remains constant during the transmission of one symbol. The value of $C_n^2$ defines how moderate or strong the atmospheric turbulence is. For such scenarios, its values ranges from $10^{-14}$ to $10^{-9}$, where $10^{-9}$ represents the strongest turbulence levels [23]. We consider $C_n^2 = 1.2 \times 10^{-11}$ m$^{-2/3}$, $2.2 \times 10^{-13}$ m$^{-2/3}$, and $3 \times 10^{-14}$ m$^{-2/3}$. Using Eqs. (3), (4), and the expression of the Rytov variance provided in Section II-A, we obtain the values of $\alpha$ and $\beta$, which are given in Table II. These values are calculated considering that the RIS structure is made in such a way that its elements and the receiver aperture have the same diameter, $D = 1$ mm. The «symengine» function in Matlab is used to generate the Meijer-G function.

First, we analyze the OP, $P_{out}$, the average channel capacity, $\overline{C}$, and the average BER, $\overline{P}_b$, for a single color (the blue), then we compare the system performance for red, green, and blue colors. Figures 2, 3, and 4 depict the OP against normalized average threshold SNR, $\gamma_{th}$, and end-to-end electrical SNR, $\overline{\gamma}$, respectively, while Figs. 5 and 6 present the ergodic channel capacity. Figures 7 and 8 give BER results for CBFSK, NBFSK, CBPSK, and DBPSK, and finally, Fig. 9 compares BER and average channel capacity results for the red, green, and blue colors for selected values of $\alpha$ and $\beta$. The analysis is carried out for $\zeta = 1.1$ and $\zeta = 6.1$, representing the pointing error level, where 1.1 is the worst case.

Figures 2, 3, and 4 respectively depict the OP across the normalized threshold SNR, $\gamma_{th}$, the normalized average electrical SNRs, $\overline{\gamma}_g$ and $\overline{\gamma}_h$, for both detection techniques. Figure 2 confirms that increasing $\gamma_{th}$ worsens the system's probability of failure for all considered turbulence levels and pointing errors. As an example, for $\zeta = 6.1$, $\alpha = 10.9537$, and $\beta = 2.9833$, adopting the HD technique, a $P_{out}$ target of $10^{-3.5}$ can be obtained for $\gamma_{th} = 0$ dB, and a $P_{out}$ target of $10^{-1}$

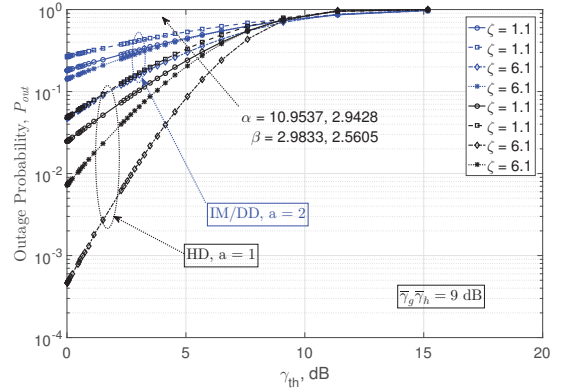

Fig. 2: OP results showing the performance of both HD and IM/DD techniques under different turbulence conditions in terms of normalized average threshold SNR, $\gamma_{th}$, for harsh ($\zeta = 1.1$) and moderate ($\zeta = 6.1$) pointing error levels. $\overline{\gamma}_g = \overline{\gamma}_h = 9$ dB.

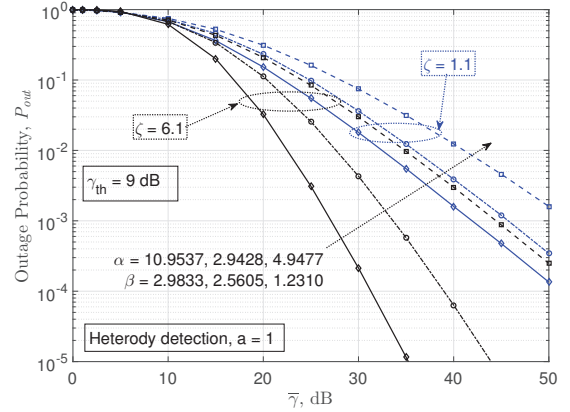

Fig. 3: OP results showing the performance of the HD technique under different turbulence conditions in terms of the average electrical SNR, $\overline{\gamma}$, for harsh ($\zeta = 1.1$) and moderate ($\zeta = 6.1$) pointing error levels. $\gamma_{th} = 9$ dB.

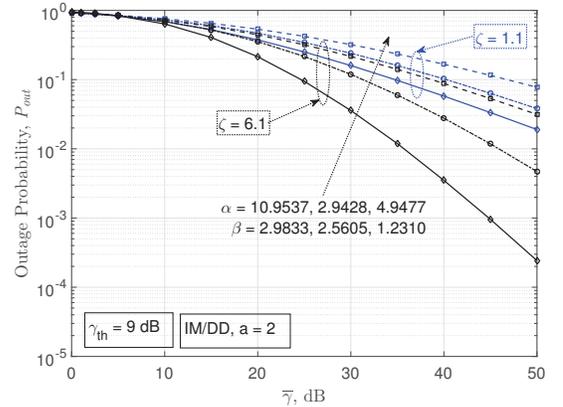

Fig. 4: OP results showing the performance of the IM/DD technique under different turbulence conditions in terms of the average electrical SNR, $\overline{\gamma}$, for harsh ($\zeta = 1.1$) and moderate ($\zeta = 6.1$) pointing error levels. $\gamma_{th} = 9$ dB.

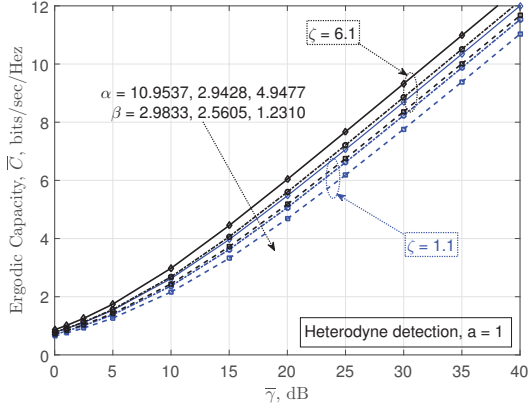

Fig. 5: Ergodic channel capacity results for the HD technique under different turbulence conditions in terms of the average electrical SNR, $\overline{\gamma}$, for harsh ($\zeta = 1.1$) and moderate ($\zeta = 6.1$) pointing error levels.

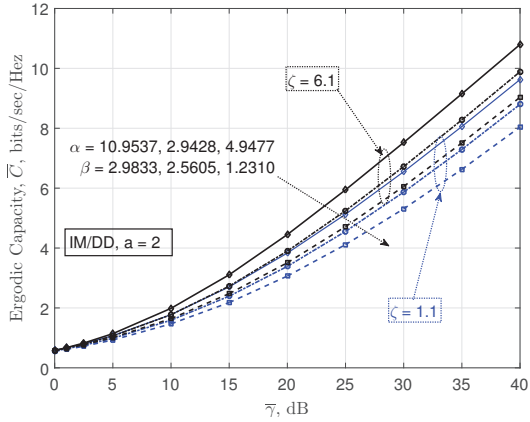

Fig. 6: Ergodic channel capacity results for the IM/DD technique under different turbulence conditions in terms of the average electrical SNR, $\overline{\gamma}$, for harsh ($\zeta = 1.1$) and moderate ($\zeta = 6.1$) pointing error levels.

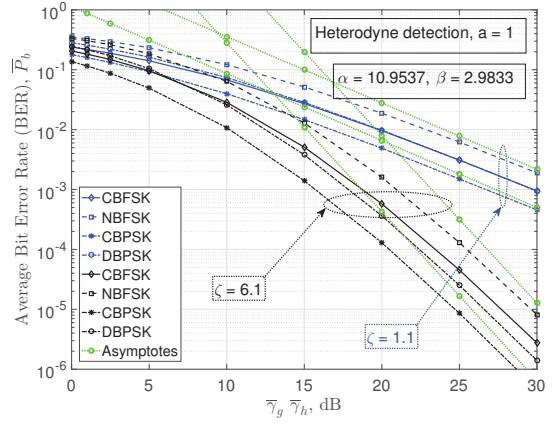

Fig. 7: Average BER of CBFSK, NBFSK, CBPSK, and DBPSK binary modulation schemes showing the HD technique's performance under different turbulence conditions for extreme ($\zeta = 1.1$) and moderate ($\zeta = 6.1$) pointing error levels.

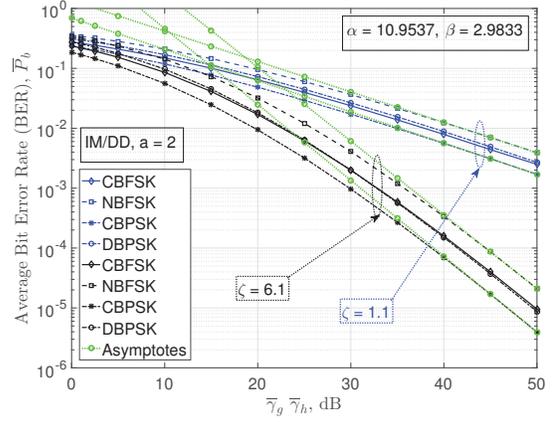

Fig. 8: Average BER of CBFSK, NBFSK, CBPSK, and DBPSK binary modulation schemes showing the IM/DD technique's performance under different turbulence conditions for extreme ($\zeta = 1.1$) and moderate ($\zeta = 6.1$) pointing error levels.

requires that $\gamma_{th}$ is set to 5.5 dB. The results in Fig. 3 show that using the HD technique for a fixed $\gamma_{th} = 9$ dB, a target $P_{out} = 10^{-3}$ is obtained at 26 dB, 33 dB, and 44 dB of $\overline{\gamma}$, respectively for atmospheric conditions characterized by $(\alpha, \beta)$ = (10.9537, 2.9833), (4.9477, 1.2310), and (2.9428, 2.5605), and a pointing error of $\zeta = 6.1$; for the same $P_{out} = 10^{-3}$ target, the system at $\zeta = 6.1$ outperforms its counterpart at $\zeta = 1.1$ of 16 dB, 13 dB, and 9 dB, respectively. This is also seen in Fig. 4, which depicts the performance of the system under the same conditions as previously, for the IM/DD technique, for the values of pointing errors $\zeta = 1.1$ and 6.1. In this case, the curves lead to the conclusion that systems employing the HD technique outperform those using the IM/DD technique of at least 12 dB in all turbulence levels. For example, for $P_{out} = 10^{-2}$, $\overline{\gamma} = 35$ dB (23 dB for the HD technique) is required if $\zeta = 6.1$, $\alpha = 10.9537$, and $\beta = 2.9833$.

Figures 5 and 6 depict the system's ergodic channel capacity for multiple atmospheric conditions considering and pointing errors, $\zeta = 1.1$ and 6.1. The curves in Fig. 5 represent results for the HD technique while the performance of the IM/DD technique is given in Fig. 6. As in the case of $P_{out}$, Figs. 5 and 6 confirm that an RIS-based nT-FSO system utilizing the HD technique outperforms the IM/DD based counterpart. They are also in agreement with the results of OP shown in Figs. 2, 3, and 4. The system offers the best ergodic channel capacity for turbulence parameters at which the OP is lower. For example, with a target $P_{out}$ of $10^{-5}$, the best result is obtained at $\overline{\gamma} = 35$ dB for a pointing error characterized by $\zeta = 6.1$ and turbulence behavior with parameters $\alpha = 10.9537$, and $\beta = 2.9833$. For the same turbulence level and pointing errors at $\overline{\gamma} = 35$ dB, the best ergodic channel capacity for the HD detection, which is about 10.2 bits/sec/Hez, is obtained for $\zeta = 6.1$, $\alpha = 10.9537$, and $\beta = 2.9833$.

Figures 7 and 8 depict the BER performance of the FSO system analyzed in this paper. They highlight the impact of pointing errors on the RIS-aided nT-FSO data transmission systems. In both figures, there is clearly a difference between the BER performance for $\zeta = 6.1$ and $\zeta = 1.1$. This is valid for all studied schemes. A CBPSK under $\zeta = 6.1$ outperforms the CBPSK under $\zeta = 1.1$. It is also clear that DBPSK slightly outperforms CBFSK, and NBFSK is the least attractive from a performance point of view. Figures 7 and 8 also illustrate the asymptotes of BER curves for selected schemes to highlight the diversity gains given by their negative slopes.

Finally, Fig. 9 compares the BER (Fig. 9a) and ergodic channel capacity (Fig. 9b) of the system assuming DBPSK for red, green, and blue colors. The values of $\alpha$ and $\beta$, obtained from the wavelengths, are $(\alpha, \beta) = (10.9537, 2.9833)$, $(12.5331, 4.6787)$, and $(13.2818, 5.7795)$, respectively. As the figures confirm, the blue light offers better performance as its corresponding BER curve is below those of green and red lights. For example, for a BER of $10^{-4}$, $\bar{\gamma} = 35$ dB is required while transmitting using a blue light. This value increases to 38 dB and 39 dB for green and red colors, respectively.

## VI. CONCLUSION

This paper has presented unified and closed-form expressions for the PDF, CDF, and MGF of a RIS-based nT-FSO link operating over G-G turbulence and pointing errors. By exploiting these expressions, the system performance has been evaluated through metrics such as the OP, ergodic channel capacity, and average BER, for selected binary schemes in terms of the Meijer-G function. The unified diversity order and coding gain for the proposed RIS-based nT-FSO system have been also derived. It has been shown, through numerical results, that RIS-assisted nT-FSO systems using the HD technique outperform those using the IM/DD technique. The influence of turbulence level and pointing errors on the transmitted signal has been highlighted. Finally, it has been shown that the transmission using the blue color outperforms those with red and green colors over a RIS-aided nT-FSO channel. In our future work, we will investigate the scenario where the line-of-sight path is non negligible and also the case of multiple RIS links.

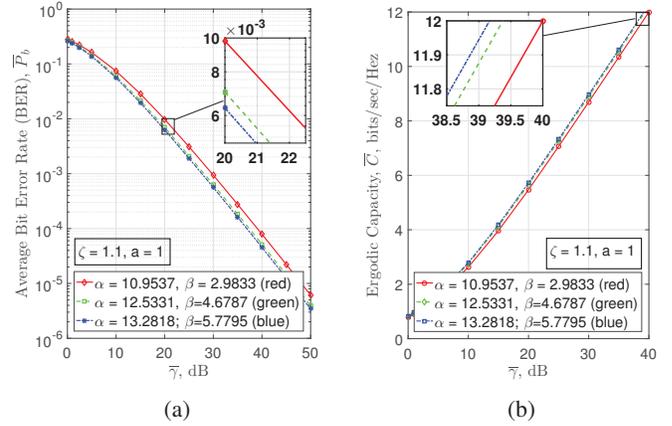

Fig. 9: Performance comparison for red, green, and blue lights: (a) average BER and (b) channel capacity.